\documentclass[twocolumn,aps,pra,amsmath,amssymb,showpacs,superscriptaddress]{revtex4}

\usepackage{ucs}
\usepackage[utf8]{inputenc} 

\usepackage{graphicx}
\usepackage{graphics}
\usepackage{latexsym}
\usepackage{amsmath}
\usepackage{amsfonts}
\usepackage{amssymb}

\begin{document}

\title{
Communication loophole in a Bell-EPR-Bohm experiment:
standard no-signaling may not always be enough to exclude local realism.
}

\author{David Rodr{\'\i}guez}
\affiliation{Departamento de F{\'\i}sica Aplicada II, Universidad de
Sevilla, E-41012 Sevilla, Spain}
\email{drodriguez@us.es}

\date{\today}


\begin{abstract}
Assuming perfect detection efficiency, we present an (indeterministic) model
for an EPR-Bohm experiment which reproduces the singlet correlations, without
contradicting Bell's original locality condition.
In this model we allow the probability distribution $\rho_{\lambda}$
of the state $\lambda$ at the source to depend parametrically on the orientation
$\xi$ of one of the measuring devices: $\rho_{\lambda}(\lambda,\xi)$.
In a Bell experiment, no-signaling between the source and each one of the devices
would seem clearly sufficient to rule such an influence;
however, not even schemes where the choice of observables takes place during the
on-flight time of the particles can prevent, in some situations, a model of this type
from violating the local bounds.
In particular, a random shift $\rho_{\lambda}(\lambda,\xi_1)
\rightarrow \rho_{\lambda}(\lambda,\xi_2) \rightarrow \ldots
\rightarrow \rho_{\lambda}(\lambda,\xi_n)$
allows the model to perform a "subensemble selection" for each of the terms involved
in the inequality (analogous to what goes on with the efficiency loophole), whenever
some correlation of those $\rho_{\lambda}$-shifts with the sequence of measurement
choices is allowed.
That correlation does not necessarily imply signaling during the photon
on-flight time.
\end{abstract}


\maketitle


\textbf{Previous note:}
\emph{
We have been made aware (much thanks to Dr. M. Hall for that) of some recent developments
on the subject \cite{DeGorre_et_al2005,Hall2010,Gisin_et_al2011,Hall2010b,Lorenzo2010}.
So far to our knowledge, results regarding our model for the singlet correlations can be
understood as a particular case of what Dr. Hall refers to as "measurement dependence",
and in general as a particular case within the broader class of models studied in most
of those references (though the formal approach is quite different).
Nevertheless, the intentions of this paper being less ambitious than most of the former,
we still think it may retain some interest as a simple, hopefully easy to grasp counterexample
to some widespread beliefs about the rigidity of the frontier between quantum and classical
behaviours.
On the other hand, it may also be useful to attract some more attention on the subject.
As a general remark, we are not so interested in "measurement dependence"
as a mathematical resource as we are in the fact that we may find a feasible and intuitive
physical counterpart for that phenomena, for instance thinking in the action of field
lines generated by the measurement devices on the source.
Besides, our approach to indeterminism (see also additional notes) and its relation
with Clauser and Horne's factorability, as little sophisticated as it may perhaps seem,
can nevertheless be, in our opinion, clarifying.
}

Following John Bell \cite{Bell64}, the results of the two distant measurements $A, B \in \{\pm 1\}$
performed in a bipartite experiment (for instance, of the EPR-Bohm type \cite{Bohm57}) can be
expressed as
\begin{eqnarray}
A &=& A(a, \lambda), \label{A}\\
B &=& B(b, \lambda), \label{B}
\end{eqnarray}
where $a,b$ are the orientations of each device, respectively, and where $\lambda \in \Lambda$,
with a probability density function $\rho_{\lambda}(\lambda)$, summarizes the state of
the pair of particles at the source.
Bell's locality condition stands, therefore, simply for the fact that $A(a, \lambda)$ does not
depend on $b$, nor does $B(b, \lambda)$ depend on $a$, either.
This (deterministic) Bell locality condition is the main hypothesis not only behind the original
Bell inequality \cite{Bell64}, but also (to our knowledge) for any other inequality
\cite{other_det_ineq}.

However, a more realistic view demands a generalization of the former expressions to the so-called
indeterministic case \cite{indeterminism}:
\begin{eqnarray}
A &=& A(a, \lambda, \omega_a), \label{A_i}\\
B &=& B(b, \lambda, \omega_b), \label{B_i}
\end{eqnarray}
where $\omega_a \in \Omega_a,\omega_b \in \Omega_b$ are other hidden variables, with density
functions $\rho_{\omega}(\omega_a,a),\rho_{\omega}(\omega_b,b)$, representing the state of the
measuring device (and therefore parametrically dependent on the corresponding orientation $a,b$).
Now, for any function $f_j(A,B)$ (subindex standing for "joint") we have
\begin{eqnarray}
&&\langle f_j(A,B) \rangle (a,b) 
= \int_{\Lambda \oplus \Omega_a \oplus \Omega_b}
d\lambda d\omega_a d\omega_b \nonumber\\
&&\quad\quad\quad\quad\quad
\rho_{(j)}(\lambda, \omega_a, \omega_b)
\cdot f_j(A(a, \lambda, \omega_a), B(b, \lambda, \omega_b))
.
\nonumber\\
\end{eqnarray}
where $\rho_{(j)}(\lambda, \omega_a, \omega_b)$ is a joint density function;
just as we did before, we simply need to invoke the intuitive idea of locality to see that we
must demand that $\lambda, \omega_a, \omega_b$ are statistically independent, i.e. their joint
probability density is factorizable: $\rho_{(j)}(\lambda, \omega_a, \omega_b) =
\rho_{\lambda}(\lambda) \rho_{\omega}(\omega_a,a) \rho_{\omega}(\omega_b,b)$.

This said, in experiments testing Bell inequalities of, for instance and again, the EPR-Bohm type
\cite{Bohm57}, a "loophole" appears (there is room for a local model reproducing the observed results)
when the two devices are not sufficiently far apart from one another, and a signal can be thus
(causally) transmitted between them, in the time that elapses between the two almost-simultaneous
measurements.
This is also known as the "locality" or "communication" loophole, and it would in principle be
excluded by the standard "no-signaling" condition (the time between the measurements is not enough
to allow for any sub-luminal transmission of information between the two locations). 
In any case, a violation of a Bell inequality is only meaningful if neither that no-signaling
nor any of the other possible additional hypothesis involved is violated too:
in the following, we will provide a model that works as a counterexample on any of the former
(bipartite) inequalities (a local model capable of reproducing the singlet correlations),
without necessarily violating the usual no-signaling condition.
This will allow us to show that, in the first place, standard no-signaling between the two
measuring devices should always be supplemented with no-signaling between the source and each
one of the devices.
But, moreover, this last restriction may be not be enough either.

Let us now consider that the probability density function for $\lambda$ is parametrically
dependent on the orientation of the nearest (to the source) device, for instance, let it be
$\rho_{\lambda}(\lambda,b)$.
Expressions (\ref{A_i})--(\ref{B_i}) remain valid, and we also demand that the three random
variables $\lambda, \omega_a,\omega_b$ keep their independence (their joint density function
remains factorizable), so now, for an arbitrary function $f_j(A,B)$ we have
(from now on, except otherwise stated, the space of integration is, naturally,
$\Lambda \oplus \Omega_a \oplus \Omega_b$):
\begin{eqnarray}
&&\langle f_j(A,B) \rangle (a,b) 
= \int_{\ }
d\lambda d\omega_a d\omega_b \times \nonumber\\
&&
\rho_{\lambda}(\lambda, b) \rho_{\omega}(\omega_a, a) \rho_{\omega}(\omega_b, b)
\cdot f_j(A(a, \lambda, \omega_a), B(b, \lambda, \omega_b)),
\nonumber\\
\label{f_j_i}
\end{eqnarray}
using the fact that $\rho_{(j)}(\lambda, \omega_a, \omega_b)$ is of course still
factorizable, as well as
\begin{eqnarray}
&&\langle f_s(A) \rangle (a) = \int_{} d\lambda d\omega_a \times
\nonumber \\
&&\quad\quad\quad\quad
\rho_{\lambda}(\lambda, b) \rho_{\omega}(\omega_a, a)
\cdot f_s( A(a, \lambda, \omega_a))
, \label{f_s_A_av}
\\
&&\langle f_s(B) \rangle (b) = \int_{} d\lambda d\omega_b \times
\nonumber \\
&&\quad\quad\quad\quad
\cdot \rho_{\lambda}(\lambda, b) \rho_{\omega}(\omega_b, b)
\cdot f_s( B(b, \lambda, \omega_b))
. \label{f_s_B_av}
\end{eqnarray}

Now, all models of this kind satisfy CH's factorability condition \cite{CH74,note_CH_fact},
as it is not difficult to prove \cite{CH_factorability}, which enables us to write
\begin{eqnarray}
&&\langle A B \rangle (a,b) =
\int d\lambda \cdot \rho_{\lambda}(\lambda, b) \times  \nonumber\\
&&\quad\quad\quad \ [ \ P(A = + 1 |a,\lambda) P(B = + 1 |b,\lambda) \nonumber\\
&&\quad\quad\quad\quad \ + \ P(A = - 1 |a,\lambda) P(B = - 1 |b,\lambda) \nonumber\\
&&\quad\quad\quad\quad\quad \ - \ P(A = + 1 |a,\lambda) P(B = - 1 |b,\lambda) \nonumber\\
&&\quad\quad\quad\quad\quad\quad \ - \ P(A = - 1 |a,\lambda) P(B = + 1 |b,\lambda) \ ]. \nonumber\\
\label{correlation}
\end{eqnarray}

At this point, we should realize that for any given
$P(A = \mu |a,\lambda)$, $\mu = \pm 1$ satisfying the axiomatic laws of probability \cite{p_axiom},
one can always find $\rho_{\omega}(\omega_a)$ consistent with that choice, and the same for
$P(B = \mu |b,\lambda)$ in relation to $\rho_{\omega}(\omega_b)$ .
So, therefore, fixing $P(A = \mu |a,\lambda),P(B = \mu |b,\lambda)$,
together with $\rho_{\lambda}(\lambda, b)$, is all we need to define our model.
Let us now make the choice
\begin{eqnarray}
P(A = \mu |a,\lambda) &=& \frac{1}{2} \left[\ 1 + \mu \cos(a-\lambda) \right], \label{pd_1} \\
P(B = \mu |b,\lambda) &=& \frac{1}{2} \left[\ 1 - \mu \cos(b-\lambda) \right], \label{pd_2}
\end{eqnarray}
with, again, $\mu \in \{\pm 1\}$,
and let us also consider that the state at the source $\lambda$ follows a probabilistic
distribution governed by the following density function
\begin{eqnarray}
\rho_{\lambda}(\lambda, b) =
\tfrac{1}{2} \left[\ \delta(\lambda - b) + \delta(\lambda - b + \pi) \ \right],
\label{source_inv}
\end{eqnarray}
where the \emph{Dirac deltas} introduce a strong, but well allowed by our hypothesis after
all, parametrical dependence of $\rho_{\lambda}$ on the direction $b$ of one of the measuring
devices.
Using the last three expressions in (\ref{correlation}) is enough to get to what we
were looking for:
\begin{eqnarray}
\langle A \cdot B \rangle (a,b) = - \cos(a-b),
\end{eqnarray}
which is precisely the quantum correlation for the singlet state.
Obviously, $\rho_{\lambda}$ is not rotationally invariant, but it will be again capable
of producing (apparent) rotationally invariant correlations from the point of view of the
experimenter \cite{apparent_rot_inv}.
Besides, this model meets the standard no-signaling requirement between the two observers, but,
surprisingly enough, still remains completely local.
A first consequence of this is that we need to include no-signaling also between each of the
pairs (observer, source), if we want to discriminate between local and non-local models.
In a highly ideal experiment where the whole correlation spectrum is analyzed, this may be
enough, but perhaps not in a real one, as we will show here.

But first, what if the parametrical dependence is for instance on $\xi \neq b$?
It is a matter of algebra to see that for
\begin{eqnarray}
\rho_{\lambda}(\lambda, \xi) =
\tfrac{1}{2} \left[\ \delta(\lambda - \xi) + \delta(\lambda - \xi + \pi) \ \right],
\label{rho_chi}
\end{eqnarray}
we obtain
\begin{eqnarray}
\langle A \cdot B \rangle (a,b) = \ - \  \cos(a-b) + \sin(a-\xi) \sin(b-\xi).
\nonumber\\
\label{corr_m}
\end{eqnarray}

We want to show to what extent a model of this kind could work in an actual experiment;
for instance, let us consider the case where the CHSH \cite{CHSH69} inequality is tested, and let
$a,a^{\prime}$ and $b,b^{\prime}$ be the two pairs of alternative orientations of the devices
at each side.
Moreover, let us adopt the procedures of Aspect's more restrictive experiment \cite{Aspect82b},
working with photons (and therefore subjected to the efficiency loophole, that we will ignore),
but still interesting for us (see \cite{interest}) because it uses (uncorrelated) post-selection
of observables for both particles, as it is described, for instance, in \cite{Aspect02}.

Now, we will suppose that $\rho_{\lambda}(\lambda,\xi)$ obeys (\ref{rho_chi}), but is randomly
shifting from one to another within a set of four possible density functions given by
$\rho_{\lambda}(\lambda,a),\rho_{\lambda}(\lambda,a^{\prime}),\rho_{\lambda}(\lambda,b)$ and
$\rho_{\lambda}(\lambda,b^{\prime})$ (therefore, $\xi \in \{a,b,a^{\prime},b^{\prime}\}$),
the four possibilities with equal probability.
These shifts are random but not completely uncorrelated with the choices of observable at the
devices.
This correlation does not necessarily need signaling (between device and source) during photon's
on-flight time, regardless of whether we use observable post-selection or not: however,
there must exist some device-source communication, though this can present some delay as well.

We will measure that correlation by a certain parameter $\Gamma$.
For instance, if we are going to measure (polarizations, projections of spin) $a,b$, then there
is a probability $\Gamma$ that $\xi$ is one of the chosen orientations (either $a$ or $b$):
a fraction $\Gamma$ of the events where we measure $a,b$ will contribute (to the observed value
of given correlation $\langle A \cdot B \rangle (a,b)$) with a value $-\cos(a-b)$.
On the other hand, for a fraction $1-\Gamma$ of those events, either $\xi = a^{\prime}$ or
$\xi = b^{\prime}$ (in principle let us suppose with equal probability), which means that
other contributions appear, modifying the overall observed correlation.
Now, using (\ref{corr_m}) we obtain that, on average over all events,
\begin{eqnarray}
&&\langle A \cdot B \rangle (a,b) = - \Gamma\cos(a-b) \nonumber\\
&&\quad\quad
- \tfrac{1}{2}(1-\Gamma)\left[\cos(a-b) - \sin(a-a^{\prime}) \sin(b-a^{\prime})\right] \nonumber\\
&&\quad\quad\quad -
\tfrac{1}{2}(1-\Gamma)\left[\cos(a-b) - \sin(a-b^{\prime}) \sin(b-b^{\prime})\right] \nonumber\\
&&\ \ = - \cos(a-b) + \tfrac{1}{2}(1-\Gamma) [\ \sin(a-a^{\prime}) \sin(b-a^{\prime}) \nonumber\\
&&\quad\quad\quad\quad\quad\quad\quad\quad\quad\quad\quad\quad\quad
+ \sin(a-b^{\prime}) \sin(b-b^{\prime}) \ ].
\nonumber\\
\end{eqnarray}
Analogous expressions for
$\langle A \cdot B \rangle (a,b^{\prime})$,
$\langle A \cdot B \rangle (a^{\prime},b)$ and
$\langle A \cdot B \rangle(a^{\prime},b^{\prime})$
(see \cite{corr_expr})
finally lead us to
\begin{eqnarray}
&&\beta_{m}(\Gamma) =
\beta_{q} + \tfrac{1}{2}(1-\Gamma) \times \nonumber\\
&&\ \ [\ \sin(a-b^{\prime}) \sin(b-b^{\prime}) + \sin(a-b) \sin(b^{\prime}-b)\nonumber\\
&&\quad + \sin(a^{\prime}-b^{\prime})\sin(b-b^{\prime}) - \sin(a^{\prime}-b) \sin(b^{\prime}-b) \ ],
\nonumber\\
\end{eqnarray}
with $\beta_{q} =
- \cos(a-b) - \cos(a-b^{\prime}) - \cos(a^{\prime}-b) + \cos(a^{\prime}-b^{\prime})$
the well known quantum mechanical prediction.

Now,
for $\Gamma=1$, shifts in $\rho_{\lambda}$ and the choices of observables $\phi_A \in \{a,a^{\prime}\}$
and $\phi_B \in \{b,b^{\prime}\}$ are completely correlated, and we obtain the quantum value;
for $1 > \Gamma > \tfrac{1}{2}$, $\rho_{\lambda}$ and $\phi_A,\phi_B$ only bear some correlation
but the model is still capable of producing a value that defies the inequality,
and for $\Gamma=0.5$, $\rho_{\lambda}$ and $\phi_A,\phi_B$ are completely uncorrelated, as a
result of which the model cannot violate the inequality.

\begin{figure}[ht!]
\includegraphics[width=1 \columnwidth,clip]{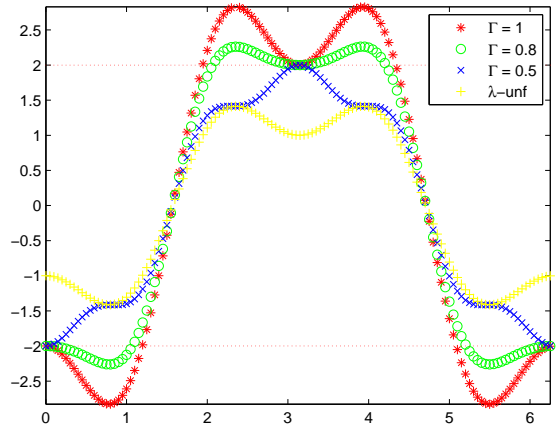}
\caption{
A maximum violation of the CHSH inequality by the quantum prediction can be achieved,
for instance, with $a=2\theta,b=\theta,a^{\prime}=0,b^{\prime}=3\theta$.
We have represented, for $\theta \in [\pi,2\pi]$:
(i) $\beta_{q} = \beta_{m}(\Gamma=1)$: * (red),
(ii) $\beta_{m}(\Gamma=0.80)$: o (green),
(iii) $\beta_{m}(\Gamma=0.50)$: x (blue),
and
(iv) the value we would obtain for a uniform distribution $\rho_{\lambda}$ (non-parametric):
$\beta_{(\lambda-unf)} = \tfrac{1}{2}\beta_{q}$: + (yellow).
This last satisfies the inequality, adding a factor of $\tfrac{1}{2}$ to the quantum
value \cite{nota_unif}.
} \label{CHSH_fig} \end{figure}

We have seen that allowing for some dependence of $\rho_{\lambda}$ on certain parameters of
the experiment ($a,b,a^{\prime},b^{\prime}$), together with some correlation (given by $\Gamma$)
of that parametric dependence on the choice of observables, makes the model still capable of
producing a value going (considerably) beyond the local bound.
Neither that dependence, nor that also necessary correlation, are completely implausible
\cite{plausability}, even assuming no-signaling during photon on-flight time, unless we design an
experiment where an exhaustive evaluation of the whole spectrum of correlations ($E: a,b \rightarrow
E(a,b)$ for all $a,b \in [0,2\pi]$) is performed, with a random, post-selected choice of
measurements for each event.

But, from the mathematical point of view... what is really happening when $\Gamma > 0.5$?
The answer is simply that, through the shift in the parametrical dependence of $\rho_{\lambda}$,
we are allowing the model to perform what is known as ``subensemble selection''
\cite{Larsson98b,note_change}: in this case the ensembles are
$\{ \Lambda_i, \rho_{\lambda} = \rho_{\lambda}(\lambda,\xi_i) \}$.
Some correlation between the choice of observables and shifts in $\rho_{\lambda}$ is needed,
nevertheless, to make it possible: with $\Gamma > 0.5$ the full ensemble
$\Lambda = \cup_{i} \Lambda_i$ is unavoidably fairly sampled.
This is the same mechanism that lays underneath the efficiency loophole, for experiments based
on data rejection (see for instance \cite{Pearle70,GM87,CRV08,CLR09}).

Nevertheless, it is still clear that the more exhaustive and restrictive the conditions of
the experiment, the less the model can get close to the quantum prediction, so margins in the
actual observed violations still play a key role.
Taking this into account, perhaps the situation with the Clauser-Horne inequality should be
studied more in detail elsewhere: there, violations are usually close to the local bound
(in part due to efficiency constraints).

Some of the first precedents of this work are \cite{Scully83} and \cite{BarutMeystre84b}.
In particular, we use the same "Malus cosine law" as Scully \cite{Scully83} for our probabilities
(\ref{pd_1})--(\ref{pd_2}) of detection (in Scully's work, probabilities of passage through an
Stern-Gerlach device).
A complete analogy with the quantum case is not achieved there, as a difference with
Barut and Meystre's work \cite{BarutMeystre84b}, where the bridge is indeed built,
although it also needs (besides some sophisticated mathematics) the introduction
of an additional (and a bit obscure) condition (some projector acting only for one of
the devices).
We also need a sort of additional assumption to make our model's and the quantum prediction meet,
but in our case its interpretation comes up as perfectly clear: we are talking of a causal
influence of one of the measuring devices on the source of the state.
This influence (taking place during the particle on-flight time or not) can find, without much
imagination, a physical, quite plausible counterpart: the effect of far field lines generated
by the device (an Stern-Gerlach, a polarizer or whatever it is).

The author thanks Dr. R. Risco-Delgado for his comments,
and acknowledges support from the Departments of Applied Physics (II and III) at Sevilla
Univ.




\begin{thebibliography}{99}
\bibitem{other_det_ineq}
For instance the Clauser-Horne-Shimony-Holt's one (CHSH) \cite{CHSH69}, or some of its main
generalizations: the Mermin inequality ($N$ parties, $2$ observables per party, \cite{Mermin90a},
based on the GHZ states \cite{GHZ89}) or that of Braunstein and Caves (two parties,
$N$ observables per party, \cite{BC89}).
\bibitem{indeterminism}
Indeterminism was firt explicitly addressed in \cite{CH74},
although it remains still perfectly compatible with the CHSH inequality \cite{CHSH69} and its
generalizations.
A necessary remark is that it is also possible to give a definition of deterministic local realism
in terms of probabilities, such as
$P(A=\mu_a) = f_A(\lambda,a), P(B=\mu_b) = f_B(\lambda,b)$ for $\mu_a,\mu_b \in \{\pm 1\}$ (or,
equivalently, $P(A=\mu_a|B=\mu_b,\lambda,a,b)=P(A=\mu_a|\lambda,a)$ and the same for $B$).

That definition is, however, always reducible to one of the kind
$A=A(\lambda^{\prime},a),B=B(\lambda^{\prime},b)$, where the hidden variable has been now redefined
from $\lambda$ to $\lambda^{\prime}=\lambda \oplus \mu$, and where $\mu$ is an extra set of variables
that also describe the state of the source.
A proof of it goes like this: let us suppose that for a given value $\lambda$ of the hidden
variable and a given set ${\cal M}$ or simultaneous (compatible) measurements, $n$ different sets
of results ${\cal R}_i$ of those measurements can occur, each one with probability $p_i$.
Then, we can define a new random variable $\mu$ taking values $\mu=\mu_i$ for each of the former
sets of results, so, naturally, $P(\mu=\mu_i|\lambda)=p_i$.
If we now build a new hidden variable $\lambda^{\prime}=\lambda \oplus \mu$, we can obtain a
completely deterministic description of the system, but this time over $\lambda^{\prime}$.

On the other hand, a hidden variable model built on $\lambda^{\prime}$ is always factorizable in
the sense of Clauser and Horne \cite{CH74}, while the one defined over  $\lambda$ may certainly not
be so.
Proof of this last fact is trivial, given that for $\lambda^{\prime}$, the probabilities involved
- $P(A|a,\lambda^{\prime}),P(B|b,\lambda^{\prime})$ - are now either unity or zero.

Therefore, the use of probabilities to define local realism does not necessarily imply indeterminism.
This indeterminism can, on the other hand, be understood in the sense of some "loss of information
in the measurement process", and described with the introduction of new variables describing the
state of the apparatus and any other factor that may locally affect the result of the measurement.
If we name this new (set of) variables $\omega_a,\omega_b$, then we will finally have
$A=A(\lambda^{\prime},a, \omega_a),B=B(\lambda^{\prime},b, \omega_b)$.
\bibitem{note_CH_fact}
The model that we propose satisfies the factorability condition assumed in Clauser-Horne's
inequality \cite{CH74}: $P(A=\mu_a,B=\mu_b|a,b,\lambda) = P(A=\mu_a|a,\lambda)\cdot P(B=\mu_b|b,\lambda)$,
for any $\mu =\pm 1$.
Though there has been some controversy on this hypothesis \cite{Selleri_et_al80,Risco02},
we will consider the problem is solved already: see \cite{indeterminism}.
\bibitem{CH_factorability}
We begin our proof by choosing, in eq. (\ref{f_s_A_av}),
$f_s(x) = 1$ for $x = \mu$ and $f_s(x)=0$ if $x \neq \mu$,
we can obtain
\begin{eqnarray}
&&P(A = \mu |a) = \nonumber\\
&&\quad\quad
\int_{A(a, \lambda, \omega_a) = \mu}
d\lambda d\omega_a \cdot \rho_{\lambda}(\lambda, b) \rho_{\omega}(\omega_a,a),
\end{eqnarray}
so
\begin{eqnarray}
&&P(A = \mu \cup \lambda|a) = \nonumber\\
&&\quad\quad\quad
\rho_{\lambda}(\lambda, b) \cdot \int_{A(a, \lambda, \omega_a) = \mu}
d\omega_a \cdot \rho_{\omega}(\omega_a,a),
\end{eqnarray}
Dividing each member by $\rho_{\lambda}$ we obtain
\begin{eqnarray}
P(A = \mu |a,\lambda) &=& \int_{A(a, \lambda, \omega_a) = \mu}
d\omega_a \cdot \rho_{\omega}(\omega_a,a),
\label{p_A}
\end{eqnarray}
that of course can be complemented with
\begin{eqnarray}
P(B = \mu |b,\lambda) &=& \int_{B(b, \lambda, \omega_b) = \mu}
d\omega_b \cdot \rho_{\omega}(\omega_b,b).
\label{p_B}
\end{eqnarray}
Now, if we also choose, for eq. (\ref{f_j_i}),
$f_j(x,y) = 1$ for $x = \mu_a, y=\mu_b$ and $f_j(x,y) = 0$ in any other case,
we can obtain the joint probabilities:
\begin{eqnarray}
&&P(A = \mu_a, B=\mu_b |a,b) = \nonumber\\
&&\quad
\int_{A= \mu_a, B = \mu_b} d\lambda d\omega_a d\omega_b \cdot
\rho_{\lambda}(\lambda, b) \rho_{\omega}(\omega_a,a) \rho_{\omega}(\omega_b,b),
\nonumber \\
\end{eqnarray}
and
\begin{eqnarray}
&&P(A = \mu_a, B=\mu_b \cup \lambda |a,b) =  \nonumber\\
&&\quad
\rho_{\lambda}(\lambda, b) \cdot \int_{A= \mu_a, B = \mu_b}
d\omega_a d\omega_b \cdot \rho_{\omega}(\omega_a,a) \rho_{\omega}(\omega_b,b),
\nonumber \\
\end{eqnarray}
which finally means that
$(A = \mu_a, B=\mu_b |a,b,\lambda) = P(A = \mu_a|a,\lambda) \cdot P(B = \mu_b|b,\lambda)$.
\bibitem{apparent_rot_inv}
When we assume (\ref{source_inv}), the rotational invariance is apparent because the paramenter
$\xi$ in $\rho_{\lambda}(\lambda,\xi)$ follows ($\xi=b$) the orientation of one of the devices
throughout its rotation.
\bibitem{corr_expr}
Using (\ref{corr_m}) we also obtain
\begin{eqnarray}
&&\langle A \cdot B \rangle (a,b^{\prime}) \nonumber\\
&&\quad
= - \cos(a-b^{\prime}) + \tfrac{1}{2}(1-\Gamma) [\ \sin(a-a^{\prime}) \sin(b^{\prime}-a^{\prime}) \nonumber\\
&&\quad\quad\quad\quad\quad\quad
+ \sin(a-b) \sin(b^{\prime}-b) \ ], \\
&&\langle A \cdot B \rangle (a^{\prime},b) \nonumber\\
&&\quad
= - \cos(a^{\prime}-b) + \tfrac{1}{2}(1-\Gamma) [\ \sin(a^{\prime}-a) \sin(b-a) \nonumber\\
&&\quad\quad\quad\quad\quad\quad
+ \sin(a^{\prime}-b^{\prime}) \sin(b-b^{\prime}) \ ], \\
&&\langle A \cdot B \rangle(a^{\prime},b^{\prime}) \nonumber\\
&&\quad
= - \cos(a^{\prime}-b^{\prime}) + \tfrac{1}{2}(1-\Gamma) [\ \sin(a^{\prime}-a) \sin(b^{\prime}-a) \nonumber\\
&&\quad\quad\quad\quad\quad\quad
+ \sin(a^{\prime}-b) \sin(b^{\prime}-b) \ ],
\end{eqnarray}
\bibitem{nota_unif}
With $\langle A B \rangle(a,b)= - \tfrac{1}{2} \cos(a-b)$ for $\rho_{\lambda}$
uniform, it is trivial that we get a
$\beta_{(\lambda-unf)} = \tfrac{1}{2}\beta_{q}$.
\bibitem{interest}
This model was initially thought for an experiment with massive particles, therefore devoid
of the efficiency loophole.
Besides, we were thinking in a EPR-Bohm scheme. Some recent experiments with atomic qubits
follow more complicated schemes \cite{RKVSIMW01, MMMOM08}, but, although the scope of our
idea can perhaps be broadened to include those other schemes, we will leave it aside for the
moment, or for elsewhere.
What we would rather do is illustrate, in some general situation, how standard procedures
do not rule out models like the one we have given.
\bibitem{plausability}
To add some plausibility to our model, we can say that, let $\tau_{flight}$ be the photon on-flight
time from the source to the device, and $\tau_{shift}$ the mean time between shifts of observable
at one of the devices, either one of the following two conditions are, for instance, sufficient to
make $\Gamma > 0.5$ possible:
(a) $\tau_{shift} > \tau_{flight}$,
(b) the probability of a shift in $\phi_A,\phi_B$ in an interval $\tau_{flight}$ is less than
$\tfrac{1}{2}$.
\end{thebibliography}
\end{document}